\documentclass[prd,12pt,superscriptaddress,nofootinbib,%
tightenlines]{revtex4}
\usepackage{mathrsfs}
\usepackage{latexsym,bm}
\usepackage{CJK}
\usepackage{graphicx}
\usepackage{indentfirst}
\usepackage{slashed}
\usepackage{amsmath}
\usepackage{amssymb}
\usepackage{color}
\usepackage{hyperref}
\usepackage{epsfig}
\hypersetup{CJKbookmarks=true}
\usepackage[titletoc]{appendix}
\usepackage{tabularx}

\newcommand{\be}{\begin{equation}} \newcommand{\ee}{\end{equation}}
\newcommand{\ba}{\begin{array}{c}} \newcommand{\ea}{\end{array}}
\newcommand{\bea}{\begin{eqnarray}} \newcommand{\eea}{\end{eqnarray}}

\begin{document}
\title{The width of the {\boldmath$\Delta$}-resonance  at two loop order\\ 
in baryon chiral perturbation theory}
\author{Jambul~Gegelia}
\affiliation{Institute for Advanced Simulation, Institut f\"ur Kernphysik
   and J\"ulich Center for Hadron Physics, Forschungszentrum J\"ulich, D-52425 J\"ulich,
Germany}
\affiliation{Tbilisi State  University,  0186 Tbilisi,
 Georgia}
\author{Ulf-G.~Mei\ss ner}
\affiliation{Helmholtz Institut f\"ur Strahlen- und Kernphysik and Bethe
   Center for Theoretical Physics, Universit\"at Bonn, D-53115 Bonn, Germany}
 \affiliation{Institute for Advanced Simulation, Institut f\"ur Kernphysik
   and J\"ulich Center for Hadron Physics, Forschungszentrum J\"ulich, D-52425 J\"ulich,
Germany}
\author{Dmitrij~Siemens}
\affiliation
{Institut f\"ur Theoretische Physik II, Ruhr-Universit\"at Bochum,  D-44780 Bochum,
 Germany}
\author{De-Liang Yao}
 \affiliation{Institute for Advanced Simulation, Institut f\"ur Kernphysik
   and J\"ulich Center for Hadron Physics, Forschungszentrum J\"ulich, D-52425 J\"ulich,
Germany}
\date{August 1, 2016}
\begin{abstract}
   We calculate the width of the delta resonance at leading two-loop order in
baryon chiral perturbation theory. This gives a correlation between the leading
pion-nucleon-delta and pion-delta couplings, which is  relevant for the
analysis of pion-nucleon scattering and other processes.


\end{abstract}
\maketitle

\newpage

   Chiral effective field theory provides  a controllable perturbative approach
of strongly interacting hadrons at low energies.  
A systematic power counting organizes  the chiral effective Lagrangian and 
observables as a perturbative series  
in the Goldstone boson sector of QCD  \cite{Weinberg:1979kz,Gasser:1983yg}.
Effective field theories (EFTs) with pions and nucleons proved to be more complicated, however,
 the problem of a consistent power counting  \cite{Gasser:1987rb} can be solved by using 
either the heavy-baryon 
approach \cite{Jenkins:1990jv,Bernard:1992qa,Bernard:1995dp}
or by choosing a suitable renormalization scheme in a manifestly Lorentz invariant formulation 
\cite{Tang:1996ca,Becher:1999he,Gegelia:1999gf,Fuchs:2003qc}.
Due to the relatively small mass difference between the nucleon and the 
$\Delta$-resonance and the strong coupling to the pion-nucleon system, the delta 
can be also included  in a systematic way in chiral EFT   (see e.g. 
Refs.~\cite{Hemmert:1997ye,Pascalutsa:2002pi,Bernard:2003xf,Pascalutsa:2006up,Hacker:2005fh}).
	A clear drawback of the low-energy EFT approach is that unlike the underlying QCD,
 the Lagrangian contains an infinite number of parameters, the low-energy constants (LECs).
However, only a finite number of them contributes to physical quantities calculated up to 
a given  order. These parameters are fixed by fitting them to 
experimental data or can be calculated on the lattice, allowing  one to  
predict other quantities. A precise determination of these LECs is an important and 
highly non-trivial task, especially when the $\Delta$-resonance is included because
there are more LECs for a given process than in the pure $\pi N$ effective Lagrangian.

In this work we calculate the width of the delta resonance in a systematic expansion 
in terms of the pion mass and the nucleon-delta mass difference (collectively denoted by $q$) 
in the framework of baryon chiral perturbation theory up-to-and-including order $q^5$, 
which includes  the leading two-loop contributions. This counting is often referred to as the
small scale expansion, see e.g. Ref.~\cite{Hemmert:1997ye}. We use the obtained results to fix a 
combination of pion-nucleon-delta couplings appearing in this expression from the experimental 
data, more precisely, we obtain  a correlation between the leading $\pi N \Delta$ and $\pi \Delta$ 
couplings.
 

The dressed propagator of the $\Delta$-resonance in $d$ space-time dimensions
can be written as
\begin{eqnarray}
-i D^{\mu\nu}(p)&=& -i \Biggl [
g^{\mu\nu}-\frac{\gamma^\mu\gamma^\nu}{d-1} -\frac{
p^\mu\gamma^\nu-\gamma^\mu p^\nu}{(d-1) m_{\Delta}^0}-\frac{d-2}{(d-1)
(m^0_{\Delta})^2}\,p^\mu p^\nu\Biggr]\, \frac{1}{p\hspace{-.35
em}/\hspace{.1em}-m_\Delta^0 -\Sigma_1 - p\hspace{-.4
em}/\hspace{.1em}\Sigma_6}\nonumber\\&+& {\rm pole \ free \
terms}\,,\label{dressedDpr}
\end{eqnarray}
where $m^0_\Delta$ is the pole mass of the delta in the chiral limit, and 
$\Sigma^{\mu\nu}$ is the self-energy of the $\Delta$-resonance. It can be parameterized as
\begin{eqnarray}
\Sigma^{\mu\nu} & = & \Sigma_1(p^2)\, g^{\mu\nu}+\Sigma_2(p^2)\,
\gamma^{\mu}\gamma^{\nu}+\Sigma_3(p^2)\,p^{\mu}\gamma^{\nu}+\Sigma_4(p^2)\,
\gamma^{\mu}p^{\nu}+\Sigma_5(p^2)\,p^{\mu}p^{\nu}
\nonumber\\
&+& \Sigma_6(p^2)\, p\hspace{-.45em}/\hspace{.1em}
g^{\mu\nu}+\Sigma_7(p^2)\,p\hspace{-.45em}/\hspace{.1em}
\gamma^{\mu}\gamma^{\nu}+\Sigma_8(p^2)\,
p\hspace{-.45em}/\hspace{.1em} p^{\mu}\gamma^{\nu}+\Sigma_9(p^2)\,
p\hspace{-.45em}/\hspace{.1em}
\gamma^{\mu}p^{\nu}+\Sigma_{10}(p^2)\,p\hspace{-.45em}/\hspace{.1em}
p^\mu p^\nu. \label{DseParametrization}
\end{eqnarray}

\begin{figure}[t]
\begin{center}
\epsfig{file=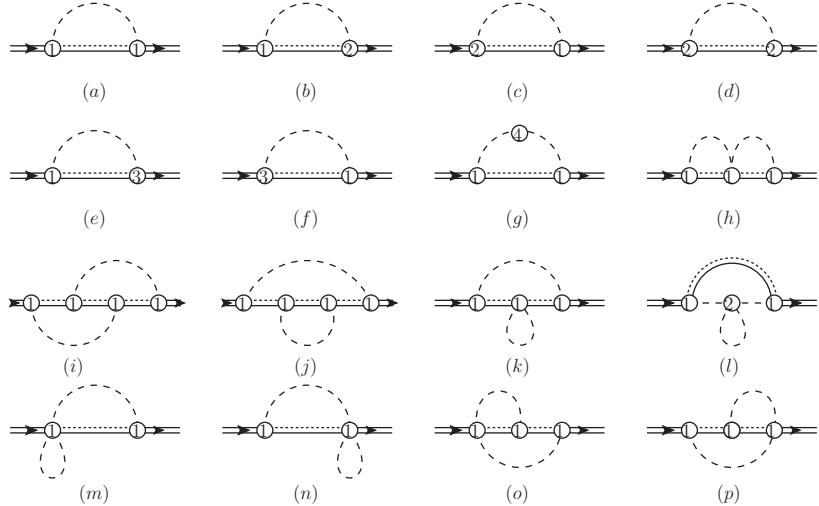,scale=0.5}
\caption{One and two-loop self-energy diagrams contributing to the width 
of the delta resonance up-to-and-including 
fifth order according to the standard power counting. The dashed and double solid lines 
represent the pions and the delta resonances, respectively. 
The double (solid-dotted) lines in the loops correspond to either nucleons or deltas. 
The numbers in the circles give the chiral orders of the vertices.}
\label{figSE2}
\end{center}
\end{figure}

\noindent
The complex pole position $z$ of the $\Delta$-propagator can be found by
solving the equation
\begin{equation}
z - m_{\Delta}^0 -\Sigma_1(z^2)-z\, \Sigma_6(z^2) \equiv z - m_{\Delta}^0 - \Sigma(z) =0\,. 
\label{poleequation}
\end{equation}
The pole mass and the width are defined by parameterizing the pole position $z$ as
\begin{equation}
z=m_\Delta-i\,\frac{\Gamma_\Delta}{2}\,.
\label{polpar}
\end{equation}

\medskip

   The one- and two-loop self-energy diagrams contributing to the 
width of the delta resonance up to order $q^5$ are shown in Fig.~\ref{figSE2}, where
the counterterm diagrams are not displayed. The underlying effective chiral Lagrangian of pions,
nucleons and the delta resonances is given in the Appendix.  For more details and the
explicit discussion of the power counting, relevant  for the current calculation of the delta 
width at leading two-loop order, we refer to Refs.~\cite{Yao:2016vbz,Gegelia:2016xcw}.

We solve Eq.~(\ref{poleequation}) perturbatively order by order in the loop expansion. 
For that  purpose  we write the self-energy as an  expansion in the number of loops
(which is equivalent to an expansion in $\hbar$)\footnote{Note that we retain the powers of
$\hbar$ for clarity here, otherwise we use natural units $\hbar=c=1$.}
\begin{equation}
\Sigma  = \hbar \,\Sigma_{(1)}+\hbar^2 \Sigma_{(2)} +{\cal O}(\hbar^3)\, ,
\label{SEpert}
\end{equation}
and obtain the following expression for the width (modulo higher order corrections)
\begin{eqnarray}
\Gamma_\Delta &=& \hbar \, 2 i \,{\rm Im} \left[\Sigma_{(1)}(m_\Delta)\right]  \nonumber\\
&+& \hbar^2 \, 2 i \, \biggl\{
{\rm Im} \left[\Sigma_{(1)}(m_\Delta)\right]{\rm Re} \left[\Sigma_{(1)}'(m_\Delta)\right]
+ {\rm Re} \left[\Sigma_{(1)}(m_\Delta)\right]{\rm Im} \left[\Sigma_{(1)}'(m_\Delta)\right] \biggr\} \nonumber\\
&+& \hbar^2 \, 2 i \, {\rm Im} \left[\Sigma_{(2)}(m_\Delta)\right] 
+{\cal O}(\hbar^3).
\label{widthpert}
\end{eqnarray}

To calculate the contributions of the one-loop self-energy diagrams to
the width, specified in the first two lines of  
Eq.~(\ref{widthpert}), we use the corresponding explicit expressions.  For the two-loop 
contribution, i.e. the terms in the third line, we use the Cutkosky cutting rules, that is we relate 
it to the corresponding decay amplitude $A_{\Delta\to\pi N}$ via
\bea
\Gamma_\Delta =\frac{\left[(m_\Delta+m_N)^2-M_\pi^2\right]\left[ \left(m_\Delta^2-m_N^2-M_\pi^2\right)^2  -4 M_\pi^2 m_N^2\right]^{3/2}}{{192}\pi m_\Delta^5}\,|A_{\Delta\to\pi N}|^2\,,
\label{widthpiN}
\eea
\medskip
where we have parameterised the amplitude for the decay $\Delta_{\mu}^i(p_i)\to \pi^a(q_a) N(p_f)$ as
\bea
{\cal A}=\bar{u}_N(p_f)\left\{A_{\Delta\to\pi N}\,q_a^\mu\xi_{ai}^{\frac{3}{2}}\right\}u_\mu(p_i)\, .
\eea

\begin{figure}[h]
\begin{center}
\epsfig{file=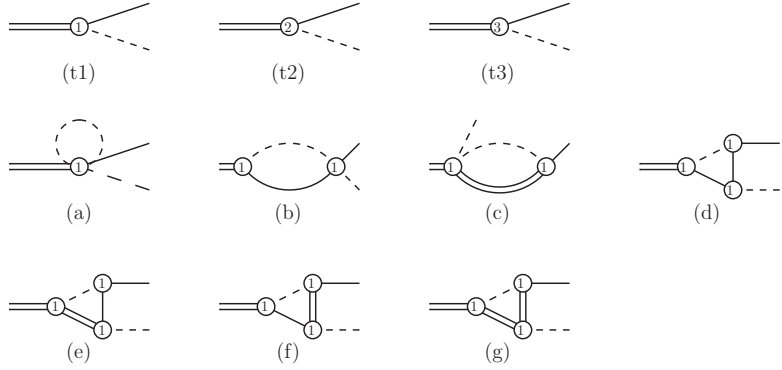,scale=0.55}
\caption{Feynman diagrams contributing to the decay $\Delta\to N\pi$ up to leading one-loop order. Dashed, solid and double 
lines represent pions, nucleons and delta resonances, respectively. Numbers in the circles mark the chiral orders of the vertices.}
\label{fig:DNPi1}
\end{center}
\end{figure}

The tree and one-loop diagrams  contributing to the $\Delta\to \pi N$  
decay  up to order $q^3$ are shown in Fig.~\ref{fig:DNPi1}. See again
Refs.~\cite{Yao:2016vbz,Gegelia:2016xcw} for the details on the power counting of the 
amplitude and the total width of the resonance.

\medskip

Calculating one- and two-loop contributions in the delta width as specified 
above we observe that by defining a linear combination of  $\pi N\Delta$ couplings 
\bea\label{eq:redefinition.h}
h_A = h - \left(b_3\Delta_{23}+b_8\,\Delta_{123}\right)
-\left(f_1\Delta_{23}+f_2\,\Delta_{123}\right)\Delta_{123}+2(2f_4-f_5)M_\pi^2\, ,
\eea
with
\bea
\Delta_{123}\equiv \frac{M_\pi^2+m_N^2-m_\Delta^2}{2 m_N}\, ,\qquad\Delta_{23}\equiv m_N-m_\Delta\, ,
\eea
modulo higher order terms, the whole explicit dependence on the couplings 
$b_3$, $b_8$, $f_1$, $f_2$, $f_4$ and $f_5$  disappears from the expression of the 
delta width. This allows us to extract  with a good accuracy the numerical value of 
$h_A$ from the experimental value of the delta width for a given value of the
leading $\pi \Delta$ coupling constant $g_1$. Such a correlation between $\pi N\Delta$ and
$\pi \Delta$ couplings exists in the large $N_C$ limit but, as far as we know, is observed
here first for the real world with $N_C=3$.

\begin{figure}[t]
\begin{center}
\epsfig{file=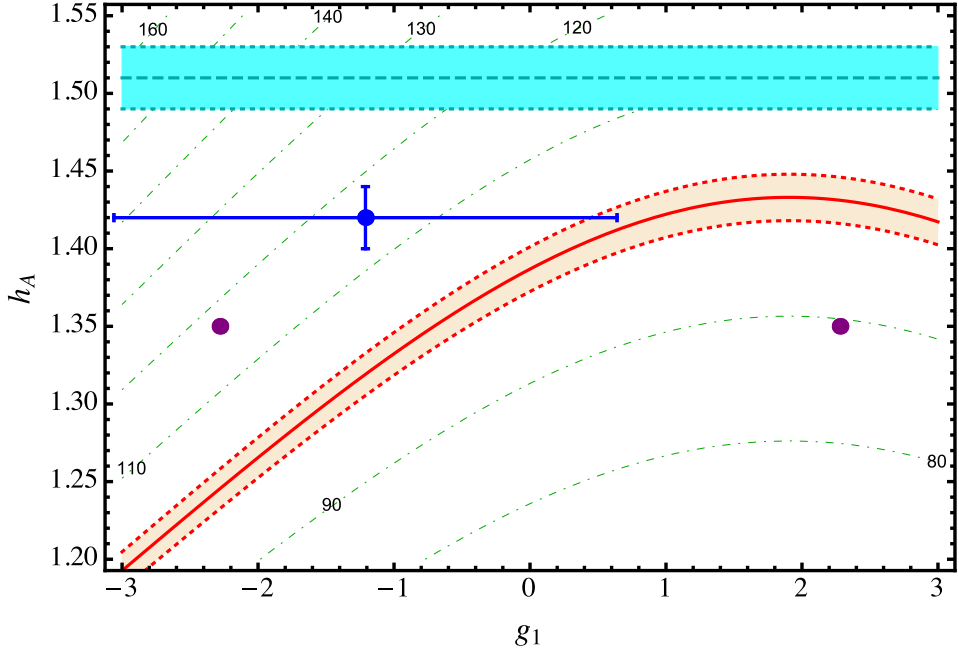,scale=0.75}
\caption{Value of the pion-nucleon-delta coupling $h_A$ as a function of 
the pion-delta coupling $g_1$ represented 
by the solid red line and the corresponding band given by the dashed red lines. 
The central line corresponds to $\Gamma_\Delta = 100$~MeV, while the band is obtained 
by varying $\Gamma_\Delta$ in the range of $98-102$~MeV  \cite{Agashe:2014kda}. 
The dot-dashed lines correspond to various values of the delta width indicated by their values 
(in MeV). For comparison, the blue dot with error bars represents the real part of the 
coupling from Ref.~\cite{Yao:2016vbz}, the purple dots stand for the values of the 
leading order pion-nucleon-delta coupling obtained in the 
large-$N_c$ limit and the horizontal dashed line with cyan band corresponds 
to the value (with error represented by the band)  from Ref.~\cite{Bernard:2012hb}.}
\label{hAgraph}
\end{center}
\end{figure}

We use the following standard 
values of the parameters \cite{Agashe:2014kda}:
$g_A = 1.27$, $M_\pi = 139\,$MeV, $m_N=939\,$MeV, $m_\Delta=1210\,$MeV, 
$F_\pi=92.2\,$MeV 
and obtain for the full decay width of the delta resonance 
\begin{eqnarray}
\Gamma_\Delta &=&  53.91\,{h}_A^2+0.87g_1^2 {h}_A^2-3.31g_1 {h}_A^2 -0.99\,{h}_A^4.
\label{GammaF}
\end{eqnarray}
Substituting $\Gamma_\Delta  = 100\pm2\ {\rm MeV}$ from the PDG in Eq.~(\ref{GammaF}), 
we extract $h_A$ as a function of $g_1$. The obtained result is plotted in Fig.~\ref{hAgraph}.  
For comparison we also show the numerical value of the $\pi N \Delta$ 
coupling from Ref.~\cite{Bernard:2012hb} (extracted at leading one-loop order and thus independent of $g_1$),  the one obtained by applying symmetry considerations in the large-$N_C$ 
limit\footnote{As the large-$N_C$ considerations do not fix the
relative sign between the two couplings, we must display two values of $g_1$ for
a given value $h_A$ here.} 
and the real 
part of the same linear combination of the couplings, as in current work, fitted to the pion-nucleon scattering phase shifts of Ref.~\cite{Yao:2016vbz}, which uses a different renormalization scheme leading to a complex valued $h_A$. Note also that Ref.~\cite{Hemmert:1997ye}
extracts 1.05 as the value of the leading order $\pi N \Delta$ coupling in the heavy baryon approach.


\medskip

  To summarize, in the current work we have  calculated the width of the delta resonance up to leading two-loop order  
in baryon chiral perturbation theory. 
Using the obtained results we fixed a combination of pion-nucleon-delta couplings, 
which also contributes in the pion nucleon scattering process, as a function of the leading pion-delta coupling.

\acknowledgments
This work was supported in part by Georgian Shota Rustaveli National
Science Foundation (grant FR/417/6-100/14) and by the DFG (CRC~110).
The work of UGM was also supported by the Chinese Academy of Sciences (CAS) President’s
International Fellowship Initiative (PIFI) (Grant No. 2015VMA076). The work of DS was supported by the Ruhr University Research
School PLUS, funded by Germany's Excellence Initiative [DFG GSC 98/3].

\appendix
\section{Effective Lagrangian}

Here, we list the relevant terms of the chiral effective Lagrangian
with pions, nucleons and deltas contributing to our calculation:
\bea
{\cal L}_{\pi N}^{(1)}&=&\bar{\Psi}_N\left\{i\slashed{D}-m+\frac{1}{2}g \,\slashed{u}\gamma^5\right\}\Psi_N\, ,\nonumber\\
{\cal L}^{(1)}_{\pi\Delta}&=&-\bar{\Psi}_{\mu}^i\xi^{\frac{3}{2}}_{ij}\left\{\left(i\slashed{D}^{jk}-m_{\Delta 0}\delta^{jk}\right)g^{\mu\nu}
-i\left(\gamma^\mu D^{\nu,jk}+\gamma^\nu D^{\mu,jk}\right) +i \gamma^\mu\slashed{D}^{jk}\gamma^\nu+m_{\Delta 0}\delta^{jk} \gamma^{\mu}\gamma^\nu\right.
\nonumber\\
 &&\left.+\frac{g_1}{2}\slashed{u}^{jk}\gamma_5g^{\mu\nu}+\frac{g_2}{2} (\gamma^\mu u^{\nu,jk}+u^{\nu,jk}\gamma^\mu)\gamma_5+\frac{g_3}{2}\gamma^\mu\slashed{u}^{jk}\gamma_5\gamma^\nu \right\}\xi^{\frac{3}{2}}_{kl}{\Psi}_\nu^l\,  ,\nonumber\\
{\cal L}^{(1)}_{\pi N\Delta}&=&h\,\bar{\Psi}_{\mu}^i\xi_{ij}^{\frac{3}{2}}\Theta^{\mu\alpha}(z_1)\ \omega_{\alpha}^j\Psi_N+ {\rm h.c.}\, ,\nonumber \\
{\cal L}^{(2)}_{\pi N\Delta}&=&\bar{\Psi}_{\mu}^i\xi_{ij}^{\frac{3}{2}}\Theta^{\mu\alpha}(z_2)
\left[i\,b_3\omega_{\alpha\beta}^j\gamma^\beta+i\,\frac{b_8}{m}\omega_{\alpha\beta}^ji\,D^\beta\right]
\Psi_N+h.c.\ ,\nonumber\\
{\cal L}^{(3)}_{\pi N\Delta}&=&\bar{\Psi}_{\mu}^i\xi_{ij}^{\frac{3}{2}}\Theta^{\mu\nu}(z_3)\left[
\frac{f_1}{m}[D_\nu,\omega_{\alpha\beta}^j]\gamma^\alpha i\,D^\beta-\frac{f_2}{2m^2}[D_\nu,\omega_{\alpha\beta}^j]
\{D^\alpha,D^\beta\}\right.\nonumber\\
&&\hspace{2.5cm}\left.+f_4\omega_\nu^j\langle\chi_+\rangle+f_5[D_\nu,i\chi_-^j]\right]\Psi_N+h.c.,
\label{LagrNDP}
\eea
where $\Psi_N$ and $\Psi_\nu$ are the isospin doublet field of the nucleon  
and the vector-spinor isovector-isospinor
Rarita-Schwinger field  of  the $\Delta$-resonance
with bare masses $m$ and $m_{\Delta 0}$, respectively. 
$\xi^{\frac{3}{2}}$ is the isospin-$3/2$ projector, 
$\omega_\alpha^i=\frac{1}{2}\,\langle\tau^i u_\alpha \rangle$ and $\Theta^{\mu\alpha}(z)=g^{\mu\alpha}
+z\gamma^\mu\gamma^\nu$. Using field redefinitions the off-shell parameters $z$  can be absorbed in 
LECs of other terms of the effective Lagrangian and therefore they can be chosen arbitrarily 
\cite{Tang:1996sq,Krebs:2009bf}. We fix the off-shell structure 
of the interactions with the delta by adopting $g_2=g_3=0$ and $z_1=z_2=z_3=0$. 
For vanishing external sources, the covariant derivatives are given by 
\begin{eqnarray}
D_\mu \Psi_N & = & \left( \partial_\mu + \Gamma_\mu 
\right) \Psi_{N}\,,  \ \ \Gamma_\mu  = 
\frac{1}{2}\,\left[u^\dagger \partial_\mu u +u
\partial_\mu u^\dagger 
\right]=\tau_k\Gamma_{\mu,k}, \nonumber\\
\left(D_\mu\Psi\right)_{\nu,i} & = &
\partial_\mu\Psi_{\nu,i}-2\,i\,\epsilon_{ijk}\Gamma_{\mu,k} \Psi_{\nu,j}+\Gamma_\mu\Psi_{\nu,i}
\,. \label{cders}
\end{eqnarray}

\end{document}